# Modeling of Whole Genomic Sequencing Implementation using System Dynamics and Game Theory

Marzieh Soltanolkottabi[a], Hadi A. Khorshidi[b], Maarten J. IJzerman[b]

[a] University of New Haven, West Haven, Connecticut, United States
[b] University of Melbourne, Melbourne, Victoria, Australia

## Abstract

Biomarker testing is a laboratory test in oncology that is used in the selection of targeted cancer treatments and helping to avoid ineffective treatments. There exist several types of biomarker tests that can be used to detect the presence of particular mutations or variation in gene expression. Whole Genome Sequencing (WGS) is a biomarker test for analyzing the entire genome. WGS can provide more comprehensive diagnostic information, but it is also more expensive than other tests. In this study, System Dynamics and Game Theoretic models are employed to evaluate scenarios, and facilitate organizational decision making regarding WGS implementation. These models evaluate the clinical and economic value of WGS as well as its affordability and accessibility. The evaluated scenarios have covered the timing of implementing WGS using time to diagnosis and total cost.

## 1. Introduction

Biomarker testing is a complex intervention to predict the risk of developing a condition and facilitate rapid and accurate diagnosis using genetic conditions that consequently leads to preventing disease, prolong life, and promoting health through early interventions [1]. Compared to single biomarker testing that may lead to a sequence of testing to diagnose the cancer, Whole Genome Sequencing (WGS) is a genomic test that sequences the whole genome with one single test. In other words, WGS is the most comprehensive test potentially allowing more biomarkers to be identified. Testing for the most prevalent biomarkers maximizes the likelihood of finding an actionable target as early as possible and minimizes the number of tests conducted. However, WGS is more costly and is mostly restricted to central facilities and/or the academic setting [2, 3, 4].

In this paper, we develop models to evaluate the process of conducting WGS against the current standard of care (SoC) diagnostics. We use System Dynamics (SD) and Game Theory to create these models. These models evaluate scenarios in terms of cost of testing and time to diagnosis.

## 2. Problem Description

WGS is a powerful test in identifying biomarkers, and if conducted early in the process of cancer treatment, it can decrease the number of tests conducted and accelerate the time to start the treatment. However, WGS is more costly in comparison with single biomarker testing and is not available in all types of hospitals. Thus, some hospitals need to send samples to other hospitals to conduct the test and this takes additional time. Although WGS can identify more biomarkers, not all patients require WGS testing, and in many cases, the SoC diagnostics is sufficient in identifying biomarkers. The problem is to identify the best time to conduct WGS testing, so the patients can start the treatment at the lowest possible cost and as early as possible.

The cancer type we are addressing in this paper is lung cancer, and the setting is what currently is in use in Netherlands [4]. In general, when a patient is diagnosed with stage IV NSCLC (Non-Small-Cell Lung Cancer), they will go through a set of biomarker testing until their diagnostic pathway is completed, then they can start the treatment. It is assumed that a patient will visit the closest hospital, and based on the type of hospital visited, the path will differ. There are three different types of hospitals, General, Teaching, and Academic. Only Academic hospitals have WGS facilities. General hospitals have the simplest types of tests, mainly testing for PD-L1, ALK, EGFR, and KRAS genes. Teaching hospitals test for ALK and PD-L1, KRAS, BRAF, EGFR, and ROS1. Academic hospitals use the same tests as teaching hospitals (SoC), but they also offer WGS testing biopsy for referred patients. It is assumed that in each hospital, these tests can be conducted in parallel. If a patient starts the process in a general hospital and their biomarkers cannot be identified, they will be referred to a teaching hospital for more tests. If the identification of biomarkers fails in the teaching hospitals, they will be referred to academic hospitals for WGS testing. If a patient is referred for WGS testing, they will be consulted to check whether they approve of the testing. Academic hospitals send biopsies to WGS facilities for WGS testing process. When WGS is conducted, the result will be sent to molecular tumor board (MTB) for interpretation. If WGS is unsuccessful, the patient will receive SoC again. In all parts of this process, referrals are based on proximity.

In this paper we are aiming to see if other scenarios such as direct referral from general hospitals to academic hospitals are better than the current scenario, and if all hospitals should follow similar scenarios.

## 3. Related Research

The common approach to analyze precision medicine technologies that treatment decisions are typically derived from a sequence of advanced biomarker testing is using cost-effectiveness analysis (CEA) for health technology assessment (HTA) and health economics. Cost-effectiveness analysis (CEA) refers to calculating the cost variation per unit of health outcomes such as prevented death or saved life-years due to some interventions. Marino, et al. [5] assess the economic impact of applying genomic testing and profiling in chemotherapy decision-making for breast cancer patients using CEA. Grosse [6] compares cost-effectiveness methods for genomic testing strategies for Lynch Syndrome in tumors from newly diagnosed patients with colorectal cancer.

There has been a debate among researchers using common CEA methods for healthcare systems and specifically for analyzing biomarker testing strategies. Some researchers believe that these common methods still fit perfectly for the purpose. On the other hand, many others have raised concerns about the ability of these methods to consider involved complexity and patient-specific details. Grosse, et al. [1] discuss the relative merits of different economic measures and methods to inform recommendations relative to genomic testing for risk of disease, including CEA and cost-benefit analysis (CBA). CBA can incorporate monetary estimates either through the direct costs of care or the indirect costs. The indirect costs can be estimated by lost economic values associated with morbidity and mortality that can be avoided by an intervention [7] or consumer willingness-to-pay to evaluate the benefits [8]. Grosse, et al. [1] conclude that CEA does not necessarily capture the full range of outcomes of genomic testing that are important for decision makers and consumers.

Neither CEA nor CBA can address the inherent complexity of healthcare systems and biomarker testing strategies. Health systems consist of multiple interdependent subsystems and processes that change dynamically and behave nonlinearly. Traditional HTA modeling methods often neglect the wider impacts that are critical for making optimal decisions and achieving desired health outcomes. CEA and CBA usually have limited usefulness and effectiveness when they are applied to complex systems [9]. Therefore, advanced methods should be applied to incorporate the dynamics and complexities of the health care systems and forecast the upstream and downstream consequences of changes in these systems. van de Ven, et al. [4] develop a Discrete Event Simulation (DES) model to evaluate the WGS implementation for NSCLC. DES is a simulation method that analyzes queuing processes and networks with focus on the utilization of resources [10]. We use SD and Game Theory to model WGS implementation for evaluating biomarker testing strategies.

## 4. Methodology

In this section, system dynamics and game theoretic models are illustrated to evaluate different scenarios. The reason to implement system dynamics model is to simulate different scenarios considering the patients flow and delay in different parts of the process. Game theoretic models are also implemented to focus more on the role of location of different facilities and the number of WGS requests on the scenario selection of hospitals.

### 4.1. System Dynamics Model

System dynamics (SD) aims to enhance the understanding of systems' complexity using causal loops for modeling, and capture nonlinearity caused by feedback processes. SD deals with dynamic behavior of systems by modeling processes over time via considering time delays in feedback loops. The art of system dynamics modeling is to represent the feedback process using causal loop diagrams (CLD) and stock and flow structures [9], [11]. In this section, we develop a conceptual model using CLD for the described problem. **Figure 1** shows the causal relationships for biomarker testing using either SoC or WGS for stage IV NSCLC patients.

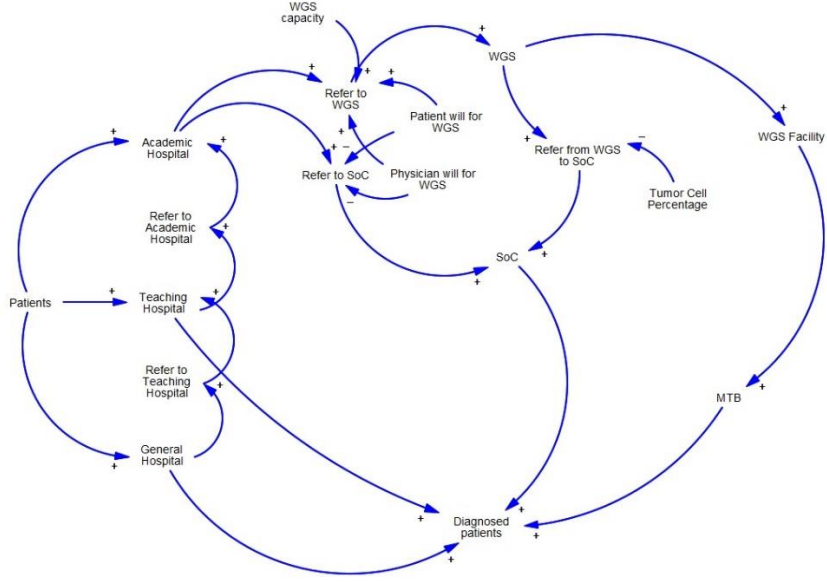

**Figure 1:** The causal loop diagram

Patients visit general, teaching, or academic hospitals based on the proximity. If the biomarker testing available in general and teaching hospitals is successful, the patient becomes a diagnosed patient and exits the system. Otherwise, the patient is referred to a more equipped hospital. In academic hospital, there is a chance to use SoC and WGS. Referral to WGS biopsy depends on willingness of the patient and physician as well as WGS capacity. Once the WGS biopsy is done, the number of the tumor cells is examined. If the percentage of tumor cells in the biopsy is low, the patient would be referred to SoC. If the biopsy is proper for WGS, the biopsy would be sent to WGS facility for further examination and preparing the thorough report. Then, the report will be sent to MTB for interpretation. Each pathway leads to cost and time to diagnosis outcomes that can be used to evaluate scenarios.

**4.2. Game Theoretic Model**

This problem is modeled as a multiplayer game with spatial interactions [12, 13], where each hospital selects one scenario from all possible scenarios to maximize its utility. The utility $U_{ij}$ of hospital i selecting scenario j consists of two parts (1) Payoff of time to start the treatment (Scenario time), and (2) Cost of scenario, which includes the cost of the tests and the traveling cost to send the samples. This utility is calculated per patient, and the payoff of time is a non-increasing function of time.

Scenario time or time to start the treatment is the summation of the time to complete the tests before WGS, WGS processing time, and traveling time. WGS processing time includes the time to complete the WGS test, and time to make decisions (MTB time), including waiting time related to the number of requests.

If a Scenario is (test series$_1$, test series$_2$, ..., test series$_i$, WGS), where test series are the set of tests that can be done in parallel, the total time of this scenario for each hospital is:

$$T_S = \sum_i P_i T_i + (1 - \sum_i P_i)T_{WGS} + traveling\ times \quad (1)$$

$$T_i = Max(t_j | j \in test\ series_i) \quad (2)$$

In which $P_i$ is the probability of starting the treatment after the i$^{th}$ test series, $t_i$ is the time to complete the i$^{th}$ test, and $T_{WGS}$ is the time to complete WGS.

WGS processing time ($T_{WGS}$) has two components, the WGS time and MTB time. The time to complete each of these depends on the number of requests in the system, and thus can be calculated using queuing theory. Assuming that the arrival of the requests to WGS hospitals and processing time are both following poison distribution, the time in the system will be $\frac{1}{\mu-\lambda}$ in which $\mu$ is the departure rate and $\lambda$ is the arrival rate. $\mu$ is a constant, but $\lambda$ depends on the number of requests. Using this philosophy, the WGS time will be as follows, in which $N_h$ is the number of WGS test requests from hospital h, H is the set of all academic hospitals, and $TN_h$ is the total number of patients at each hospital.

$$WGS\ Time = \frac{1}{WGS\ facility\ processing\ capacity - \sum_{h \in H} N_h} \quad (3)$$

$$N_h = (1 - \sum_i P_i) TN_h \quad (4)$$

MTB time will be calculated in the same manner.
In this model, each hospital selects the scenario from the possible scenarios that maximizes its utility.

## 5. Results

In this section, we implement the proposed SD and game theory models, and examine two scenarios. Scenario 1 represents the current flow of the system which is described in section 2 and visualized in **Figure 1** using CLD. Scenario 2 is for the situation that the unsuccessful cases in general hospital are directly referred to an academic hospital, so that patients who went to general hospital initially would have WGS testing with a higher chance.

**Table 1** presents the success probability which is the probability of finding the mutation (prevalence) and the probability of performing the test without technical failures, and the cost of each test in euros. The estimated values have been derived from [4]. The success probability of testing, their duration and cost for each hospital are presented in **Table 2**. The annual expected patient population is 5313, from them 60% visit general hospitals, 29% visit teaching hospitals and 11% visit academic hospitals. The cost for WGS test per patient is 2925.25 euros, and average time for WGS is 14 days. Both patient and physician's willingness for WGS test is 90%, and the probability of having sufficient Tumor cell in WGS biopsy is 0.66. WGS capacity is 1600 biopsies per year.

**Table 1:** Details for biomarker tests

| Biomarkers | Success probability | Cost per patient (euros) |
|---|---|---|
| **PD-L1** | 0.22 | 93.74 |
| **ALK** | 0.03 | 101.88 |
| **EGFR** | 0.1 | 71.19 |
| **KRAS** | 0.38 | 67.33 |
| **BRAF** | 0.02 | 100 |
| **ROS1** | 0.01 | 101.88 |

**Table 2:** Estimated values for hospitals

| Hospital | Testing success probability | Turnaround time (days) | Cost per patient (euro) |
|---|---|---|---|
| **General** | 0.58 | 15 | 334.14 |
| **Teaching** | 0.6 | 15 | 536.02 |
| **Academic (SoC)** | 0.6 | 18 | 536.02 |

### 5.1. System Dynamics Model

We develop a stock and flow model using CLD to simulate the system and the scenarios using SD. **Figure 2** shows the simulation model. This model has been run under each scenario for one year. The number of diagnosed patients in each day and total daily cost of testing have been used to compare two scenarios. The results can be seen in **Figure 3**. Scenario 2 can increase the number of diagnosed patients, so that the time to diagnosis would decline. However, the total cost for scenario 2 is higher than scenarios 1. In this example, once the trends are stabilized, the daily number of diagnosed patients for scenario 2 is 33% higher than scenario 1, while the total cost of scenario 2 is 20% higher than scenario 1.

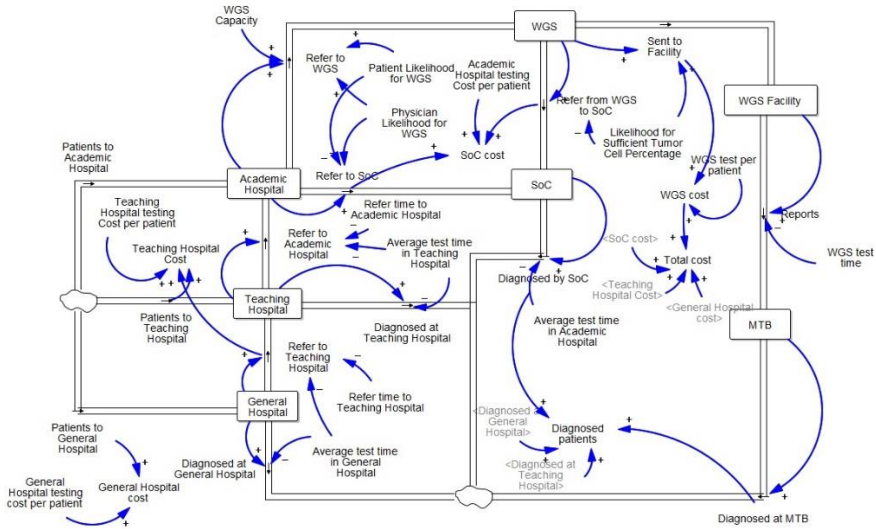

**Figure 2:** The stock and flow model

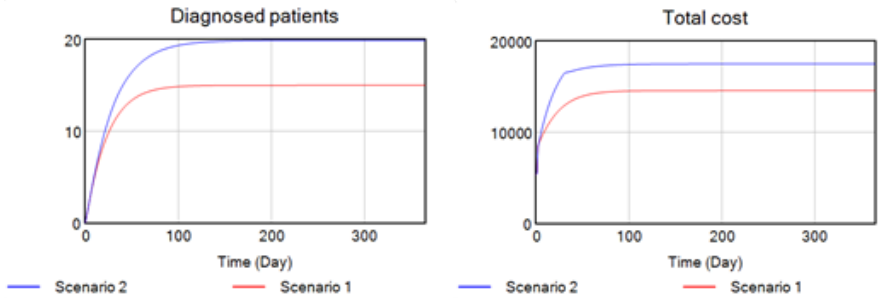

**Figure 3:** Results using the stock and flow model

### 5.2. Game Theoretic Model

In this part, a model is built to calculate the utility of different hospitals. 5o hospitals are randomly generated in an area of 100 by 100 Km, from which 60% are general hospitals, 29% are teaching hospitals, and 11% are academic hospitals. It is assumed that there is only one WGS facility available in the area, and patients are randomly allocated to different hospitals. The distance traveled in one hour is assumed to be 80 Km, and the cost of traveling one unit of distance is assumed to be 1 euro. The Capacity of WGS facility and MTB are 1600 and 9720 per year respectively. The utility is calculated using the logic described in **Section 4.2**.

To compare the illustrated scenarios, the total utility of the hospitals under each scenario is calculated. Based on our mathematical formulation, the utility of the hospitals depends on the value of the time in starting the treatment. **Figure 4** shows the total utility under each scenario for different values of time delay cost. It can be seen that Scenario 1 has higher total utility when the value of time is low, as the value of time is starting the treatment increases, Scenario 2 becomes more worthwhile.

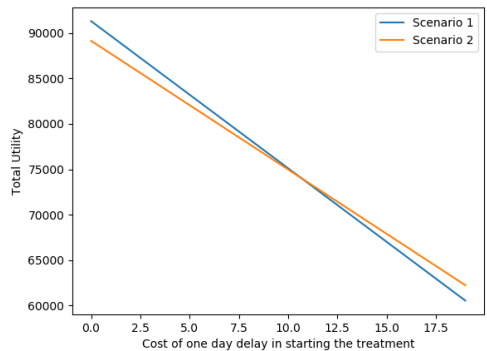

**Figure 4:** Total utility under different scenarios for different values of time

## 6. Conclusions

In this paper, we propose two models for implementing whole genomic sequencing (WGS) and biomarker testing for cancer diagnosis using system dynamics (SD) and game theory. The models are developed for patients in their stage IV NSCLC (Non-Small-Cell Lung Cancer) in Netherlands. Two scenarios have been examined. Scenario 1 represents the current procedure, while scenario 2 would expedite having WGS by referring patients from general hospitals to academic hospitals where host WGS biopsy. The SD results show scenario 2 increases the daily number of diagnosed patients while it has more cost. The Game theory model shows the importance of the value of time in decision making. Scenario 1 is a better choice if the value of starting the treatment earlier is small, while scenario 2 is better when time has a higher value. It can be seen from both models that although scenario 1 seems a less costly choice, scenario 2 is a more efficient choice if not just the monetary profit of scenarios is taken into account. Future directions of research include examinations of various scenarios, as well as the respective optimal strategy selection for different hospitals.

## Acknowledgements

We acknowledge the support from the project "Illumina - UoM Health Economics and Genomics platform" for rapid translation of genomics into health services. We also would like to acknowledge Erik Koffijberg and Michiel van de Ven for sharing their work on Discrete Event Simulation model.